\documentclass[superscriptaddress,secnumarabic,prd,aps,showpacs,nofootinbib,showkeywords,eqsecnum,preprint]{revtex4-1}

\usepackage{graphicx,color,amsmath,amsxtra}
\usepackage{epsf}
\usepackage{amssymb}
\usepackage{enumerate}
\usepackage{hhline}
\usepackage{array}
\usepackage{tabularx}
\usepackage[unicode]{hyperref}
\usepackage{graphicx}                
\usepackage{epstopdf}

\begin{document}
\pagestyle{myheadings}

\title{A note on the area of event horizon of Kleinian black hole}
\author{Tuan Q. Do }
\email{tuan.doquoc@phenikaa-uni.edu.vn}
\affiliation{Phenikaa Institute for Advanced Study, Phenikaa University, Hanoi 12116, Vietnam}
\affiliation{Faculty of Basic Sciences, Phenikaa University, Hanoi 12116, Vietnam}
\date{\today} 
\begin{abstract}
We point out that the area of event horizon of Kleinian black hole is infinite due to the fact that its event horizon is not a sphere but a hyperboloid. Therefore, the usual interpretations of Schwarzschild black hole might not be applicable to the Kleinian black hole. 
\end{abstract}

\maketitle
\section{Introduction} \label{intro}
Black holes have been one of the most attractive theoretical topics not only in the framework of Einstein's theory of general relativity but also in the framework of theoretical physics. Especially, recent observational evidences of the existence of black holes make this topic hotter than before. Besides the well-known black holes, many new or exotic black holes have been constructed by many people via various scenarios. Recently, the so-called  Kleinian black hole has been constructed explicitly by Easson and Pezzelle in Ref. \cite{Easson:2023ytf}. This paper is a follow-up study of a recent paper in Ref. \cite{Crawley:2021auj}. In their paper \cite{Easson:2023ytf}, Easson and Pezzelle have proved that there is a unique solution in split-signature spacetimes with Kleinian $SO(2,1)$ spherical symmetry. In particular, a Kleinian black hole in vacuum has the following metric  \cite{Easson:2023ytf},
\begin{equation} \label{Kleinian-bh}
ds^2 = \left(1+\frac{A}{r}\right)dt^2+\left(1+\frac{A}{r}\right)^{-1} dr^2 -r^2 d\theta^2 -r^2 \sinh^2 \theta d\phi^2.
\end{equation}
Interestingly, through the analytic continuation, $t\to it$ and $\theta \to i\theta$, along with the setting $A=-2m$, this metric can be reduced to the well-known Schwarzschild black hole,
\begin{equation} \label{Sch-bh}
ds^2 = -\left(1-\frac{2m}{r}\right)dt^2+\left(1-\frac{2m}{r}\right)^{-1} dr^2 +r^2 d\theta^2 +r^2 \sin^2 \theta d\phi^2.
\end{equation}
Ones might therefore state that the Kleinian metric is a complexified version of the Schwarzschild one \cite{Easson:2023ytf}. 
It has been shown in Ref.  \cite{Easson:2023ytf} that the Kleinian black hole metric \eqref{Kleinian-bh} will reduce to the flat metric in the Kleinian signature,
\begin{equation}
ds^2 = dt^2-dx^2-dy^2+dz^2.
\end{equation}
It turns out that the Kleinian signature has been shown to be useful to studies of quantum field theory, quantum gravity, and even black holes, e.g., see Refs. \cite{Easson:2023ytf,Crawley:2021auj,Heckman:2022peq,Arkani-Hamed:2019ymq,Barrett:1993yn,Crawley:2023brz,Easson:2023dbk,Adamo:2023fbj,Guevara:2023wlr,Desai:2024fgr}. 

On the other hand, the Schwarzschild black hole metric \eqref{Sch-bh} will reduce to the flat (Minkowski) metric in the Lorentzian signature associated with $SO(3)$ spherical symmetry,
\begin{equation} \label{Lorentzian}
ds^2 =-dt^2+dx^2+dy^2+dz^2.
\end{equation}
In the next section, we are going to calculate the corresponding area of event horizon of Kleinian black hole. As a result, we will point out that this area turns out to be infinite, in contrast to that of the well-known Schwarzschild black hole. 

This paper will be organized as follows: (i) A brief introduction of the Kleinian black hole has been written in Sec. \ref{intro}. (ii) Area of event horizon of the Kleinian black hole will be discussed in Sec. \ref{sec2}. (iii) Concluding and further remarks will be shown in Sec. \ref{final}. In addition, some calculations will be presented in the Appendix.
\section{Area of event horizon of the Kleinian black hole} \label{sec2}
\begin{figure}[hbtp] 
 \begin{center}
 	\includegraphics[scale=0.7]{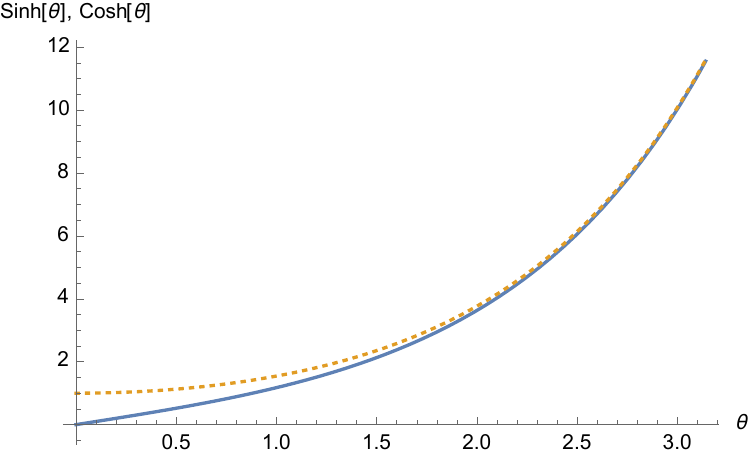}\\
 	\caption{Behavior of $\sinh\theta$ (solid curve) and $\cosh\theta$ (dotted curve) for $0\leq \theta \leq \pi$.}
 	\label{fig1}
	\end{center}
 \end{figure}

\begin{figure}[hbtp] 
 \begin{center}
 	\includegraphics[scale=0.6]{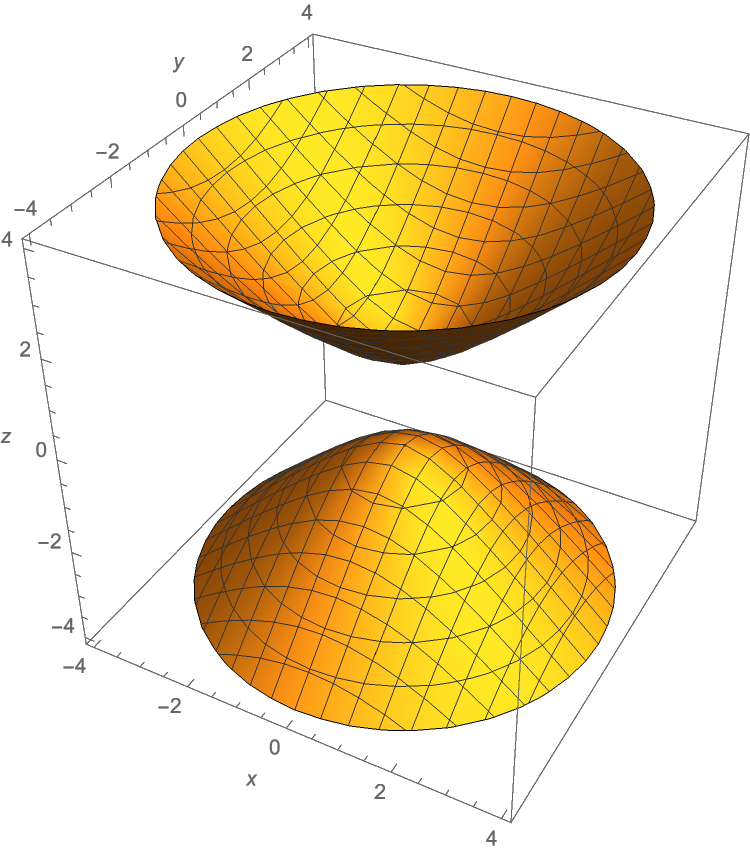}\\
 	\caption{Illustration of two-sheeted hyperboloid corresponding to an equation $-x^2-y^2+z^2 = 1$.}
 	\label{fig2}
	\end{center}
 \end{figure}

It is noted that the metric \eqref{Kleinian-bh} has been written in the coordinates $(r,\theta,\phi)$ due to  the following coordinate transformations,
\begin{align}
x &=r \cos\phi \sinh \theta,\\
y&= r\sin\phi \sinh \theta,\\
z&= r\cosh \theta,
\end{align}
with $ 0\le r$, $0\leq \phi \leq 2\pi$, and $0\leq \theta <+ \infty$ \cite{Easson:2023ytf,Crawley:2021auj}. It turns out that $\sinh \theta$ and $\cosh \theta$ are positive definite for $0\leq \theta <+ \infty$ (see Fig. \ref{fig1} for their graphs). This indicates that $z\ge 0$ for $0\leq \theta  <+ \infty$. Therefore, the Kleinian black hole will correspond to a positive sheet of two-sheeted hyperboloid described by the corresponding equation \cite{Weisstein} (see Fig. \ref{fig2} for an illustration), 
\begin{equation}
-x^2-y^2+z^2 =r^2.
\end{equation}

 
 It should be noted again that the Schwarzschild black hole can be seen, by an observer located outside the event horizon, as a spherical object in three dimensional space. This is due to the fact that the Schwarzschild black hole is static and written in the Lorentzian signature. Its event horizon is therefore is a sphere having a radius $r=r_h$. Therefore, the corresponding area of event horizon of the Schwarzschild black hole  is defined as
 \begin{equation}
A_{\rm Sch} = 4\pi r_h^2.
\end{equation}
 
\begin{figure}[hbtp] 
 \begin{center}
 	\includegraphics[scale=0.6]{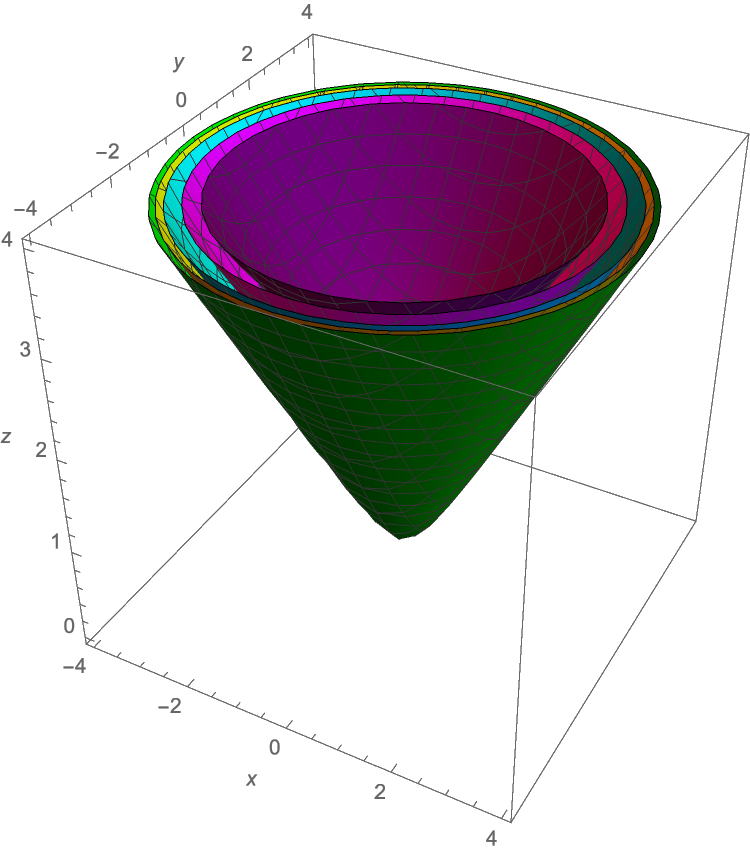}\\
 	\caption{Three dimensional illustration of the Kleinian black hole corresponding the hyperboloid equation $-x^2-y^2+z^2 =r^2$. Green, yellow, cyan, magenta, and purple surfaces correspond to $r^2=0.2, ~ 1, ~2, ~4$, and $6$, respectively.}
 	\label{fig3}
	\end{center}
 \end{figure}

 One might ask if the area of event horizon of the Kleinian black hole is equal to that of the Schwarzschild black hole, provided that $A=-2m$. It appears that the event horizon of the Kleinian black hole \eqref{Kleinian-bh} is also located at $r_h=2m$, similar to that of the Schwarzschild black hole \eqref{Sch-bh}. As a result, we are able to define the corresponding area of event horizon of the Kleinian black hole as follows (see Appendix for detailed calculations),
 \begin{equation}
 A_{\rm Kle} = r_h ^2 \int_0^{2\pi} d\phi \int_0^{+\infty} \sinh \theta \sqrt{1+2\sinh^2 \theta} d\theta =2\pi  r_h ^2 I,
  \end{equation}
 with 
 \begin{equation}
 I = \int_0^{+\infty} \sinh \theta \sqrt{1+2\sinh^2 \theta} d\theta.
 \end{equation}
 Unfortunately, this integral is not convergent. This means that 
 \begin{equation}
 A_{\rm Kle} \gg  A_{\rm Sch}.
 \end{equation}
   To see the geometry of the Kleinian black hole, we plot the hyperboloid equation  $-x^2-y^2+z^2 =r^2$ with $r^2 =0.2$, $r^2=1$ (we set $r_h=1$ for convenience), $r^2=2$, $r^2=4$, and $r^2=6$. It appears that for $r > r_h \equiv 1$ the corresponding surface is located within a region surrounded by the event surface. Of course, for $r<r_h \equiv 1$, the corresponding surface is larger than the event horizon. This feature of the Kleinian black hole is indeed contrast to that of the Schwarzschild one (see Fig. \ref{fig3} for more details).
 
Before ending this section, we would like to discuss an alternative analytic continuation, from the Lorentzian metric shown in Eq. \eqref{Lorentzian} to a Kleinian one, given in the Appendix A of Ref. \cite{Easson:2023ytf}. As a result, this continuation is done by taking a minimal complexification on $z \to iz$ alone, by which the corresponding flat Kleinian metric turns out to be
 \begin{equation} \label{Kleinian-new}
 ds^2 =-dt^2 -dz^2 +dx^2+dy^2.
 \end{equation} 
Therefore, the metric of the Kleinian black hole in vacuum shown in Eq. \eqref{Kleinian-bh} must be modified accordingly in order to be consistent with the alternative Kleinian metric described by Eq. \eqref{Kleinian-new}. In particular, the metric of the Kleinian black hole in this case takes the corresponding  form \cite{Easson:2023ytf},
\begin{equation}
ds^2 = - \left(1+\frac{A}{r}\right)dt^2+\left(1+\frac{A}{r}\right)^{-1} dr^2 -r^2 d\theta^2 +r^2 \cosh^2 \theta d\phi^2,
\end{equation}
thanks to the following coordinate transformations,
\begin{align}
x &=r \cos\phi \cosh \theta,\\
y&= r\sin\phi \cosh \theta,\\
z&= r\sinh \theta.
\end{align}
After some simple algebra, we arrive at the corresponding area of event horizon of the alternative Kleinian black hole metric given by
\begin{equation}
 A_{\rm Kle}^{\rm alt} = r_h ^2 \int_0^{2\pi} d\phi \int_0^{+\infty} \cosh \theta \sqrt{1+2\sinh^2 \theta} d\theta =2\pi  r_h ^2 I^{\rm alt},
  \end{equation}
  with 
 \begin{equation}
 I^{\rm alt} = \int_0^{+\infty} \cosh \theta \sqrt{1+2\sinh^2 \theta} d\theta = +\infty.
 \end{equation}
 This result does confirm the fact that the area of event horizon of the Kleinian black hole is always infinite, regardless of the types of analytic continuation considered above. 
 \section{Conclusions and further remarks} \label{final}
The results presented above indicate that the usual interpretations of the Schwarzschild black hole might not be applicable to the Kleinian black hole. In addition, the Kleinian black hole might just be a mathematical solution associated with the Kleinian signature useful to the studies of quantum field theory, quantum gravity, and even black holes, e.g., see Refs. \cite{Heckman:2022peq,Arkani-Hamed:2019ymq,Barrett:1993yn,Crawley:2023brz,Easson:2023dbk,Adamo:2023fbj,Guevara:2023wlr}. In other words, the Kleinian black hole could  not be a physical object but rather a tool to use in computation. See also Refs. \cite{Easson:2023ytf,Crawley:2021auj} for detailed discussions on the importance of Kleinian signature. Of course, one might claim that the infinite area of event horizon of the Kleinian black hole is purely an artifact of coordinate transformations because of the unbounded angular coordinate $\theta$. On the other hand, one might think that the so-called area of the surface defined by $r=r_h$ of the Kleinian black hole does not resemble the area of event horizon of the corresponding Schwarzschild  black hole having the Lorentzian signature. However, no matter it is mathematical and/or physical black holes, thermodynamics quantities of the Kleinian black hole, e.g., its temperature and entropy, need further explorations, by which we can have more information to compare it with its counterpart,  the Schwarzschild black hole. 
\begin{acknowledgments}
The author would like to thank an anonymous referee very much for his/her useful comments and suggestions. The author would also like to thank Prof. W. F. Kao, Prof. Damien A. Easson, and Dr. Rahul Kumar Walia very much their useful comments. Additionally, the author appreciates Prof. Gabriel Herczeg  for his correspondence. 
\end{acknowledgments} 
\appendix
\section{Calculations of an event horizon area of the Kleinian black hole}
In this Appendix, we would like to present a derivation for an event horizon area of the Kleinian black hole. Firstly, we note that the usual formalism for defining an event horizon area of the Schwarzschild black hole (and other spherical black holes such as the Kerr one) \cite{Carroll},
\begin{equation} \label{formalism-1}
A_{\rm Sch} = \iint \sqrt{|\gamma|} d\phi d\theta = r_h^2 \int_0^{2\pi} d\phi \int_0^\pi \sin\theta d\theta = 4\pi r_h^2,
\end{equation}
where $|\gamma|$ is the determinant of the induced metric $\gamma_{ij}$ on the event horizon with $r=r_h$, $dt=0$, and $dr=0$. More specifically, 
\begin{equation}
\gamma_{ij}dx^i dx^j = r_h^2 d\theta^2 +r_h^2 \sin^2\theta d\phi^2.
\end{equation}
Interestingly, this result can also be obtained using the surface integral for sphere with the radius $r=r_h$. Indeed, a parametric representation of this sphere can be defined to be
\begin{equation}
{\bf r} (\phi,\theta)= r_h \left[ \cos \phi \sin \theta, \sin\phi \sin\theta,\cos\theta \right],
\end{equation}
with $0\leq \phi \leq 2\pi$ and $0\leq \theta \leq \pi$. Then, the area of this sphere is given by
\begin{equation}
A_{\rm Sch} = \iint dA = \iint |{\bf r}_\phi \times {\bf r}_\theta|d\theta d\phi = r_h^2 \int_0^{2\pi} d\phi \int_0^\pi \sin \theta d\theta =4\pi r_h^2.
\end{equation}
Here ${\bf r}_\phi \equiv \partial {\bf r}/\partial \phi$ and ${\bf r}_\theta \equiv \partial {\bf r}/\partial \theta$ as the tangent vectors of the sphere, which span the tangent plane of the sphere.

Now, we move on to the Kleinian black hole, whose event horizon is not a sphere at all. Instead, it is a hyperboloid. If we use the formalism \eqref{formalism-1} to calculate the area of event horizon of the Kleinian black hole, we will obtain 
\begin{equation} \label{formalism-2}
A_{\rm Kle} = \iint \sqrt{|\gamma|} d\phi d\theta = r_h^2  \int_0^\pi d\phi \int_0^{+\infty} \sinh \theta d\theta = + \infty.
\end{equation}
On the other hand, we are able to define the corresponding area of event horizon of the Kleinian black hole using the surface integral technique such as
\begin{equation} \label{formalism-3}
A_{\rm Kle}= r_h^2  \int_0^\pi d\phi \int_0^{+\infty} \sinh \theta \sqrt{\sinh^2 \theta +\cosh^2 \theta} d\theta = + \infty.
\end{equation}
It should be noted that this hyperboloid is parameterized as 
\begin{equation}
{\bf r} (\phi,\theta)= r_h \left[ \cos \phi \sinh \theta, \sin\phi \sinh\theta,\cosh\theta \right],
\end{equation}
with $0\leq \phi \leq 2\pi$ and $0\leq \theta <+\infty$. It is noted that there is a gap between two integrals shown in Eqs. \eqref{formalism-2} and \eqref{formalism-3} due to the existence of $\sqrt{\sinh^2 \theta +\cosh^2 \theta}$. This indicates a hint that the formalism \eqref{formalism-2} might not be applicable to the Kleinian black hole having the hyperboloid horizon.

\end{document}